\documentclass[12pt]{article}

\usepackage{mathrsfs}

\usepackage{amsmath,amssymb}
\usepackage{amsmath}
   \usepackage{amscd}
   \usepackage{amsfonts}
   \usepackage{amssymb}
   \usepackage{amsthm}
   \usepackage{epsfig}

\def\be{\begin{equation}}
\def\ee{\end{equation}}
\def\ba{\begin{eqnarray}}
\def\ea{\end{eqnarray}}

\def\ket[#1]{\left|#1\right>}
\def\bra[#1]{\left<#1\right|}

\def\tA{{\tilde A}}
\def\tB{{\tilde B}}
\def\tC{{\tilde C}}
\def\tD{{\tilde D}}

\def\ket[#1]{\left|#1\right>}
\def\bra[#1]{\left<#1\right|}

\parskip=\medskipamount

\input epsf
\usepackage{epsfig,color,latexsym}

\renewcommand{\theequation}{\thesection.\arabic{equation}}

\begin{document}
\renewcommand{\theequation}{\thesection.\arabic{equation}}

%
%
\setcounter{footnote}{0}
\def\thefootnote{\arabic{footnote}}

\renewcommand{\theequation}{\thesection.\arabic{equation}}

\thispagestyle{empty}
\vspace{4em}

\begin{center}
\textbf{\Large\mathversion{bold} Supermembrane actions for Gaiotto-Maldacena backgrounds}

\vspace{2em}

\textrm{\large Bogdan Stefa\'nski, jr.} 

\vspace{2em}

\vspace{1em}
\begingroup\itshape
Centre for Mathematical Science, City University London, \\ Northampton Square,  London  EC1V 0HB , UK
\par\endgroup

\vspace{1em}

\texttt{Bogdan.Stefanski.1@city.ac.uk}


\end{center}

\vspace{6em}

\begin{abstract}\noindent
We write down the supermembrane actions for M-theory backgrounds dual to general ${\cal N}=2$ four-dimensional superconformal field theories. The actions are given to all orders in fermions and are in a particular $\kappa$-gauge. When an extra $U(1)$ isometry is present our actions reduce to $\kappa$-gauge fixed Green-Schwarz actions for the corresponding Type IIA backgrounds.
\end{abstract}

\newpage

\section{Introduction}
\setcounter{equation}{0}

${\cal N}=2$ supersymmetric Yang-Mills (SYM) theories have played an important role in understanding the non-perturbative behaviour of gauge theories in four-dimensions~\cite{sw}.  Gaiotto~\cite{g1} has provided a construction of a very large family of four-dimensional ${\cal N}=2$ superconformal field theories(SCFTs) using Riemann surfaces with suitable punctures. Dual M-theory space-times for these gauge theories were written down by Gaiotto and Maldacena (GM)~\cite{gm}. These solutions are closely related to the solutions investigated in~\cite{llm} and~\cite{lm} and were found using techniques first applied to ${\cal N}=1$ backgrounds in~\cite{Gauntlett:2004zh}. The GM solution contains an $AdS_5$ factor, an $S^2$ and an $S^1$ which geometrically encode the bosonic symmetries of the N=2 superconformal algebra $SU(2,2|2)$. The three factors have different relative warpings which are functions of the three remaining coordinates $y\,,\,x_1\,,\,x_2$, with the $S^1$ being in addition fibred over these three coordinates. The full solution is determined by a single function $D(y,x_1,x_2)$ which satisfies the three dimensional Toda equation
\begin{equation}
\partial_{x_1}^2 D+\partial_{x_2}^2 D+\partial_{y}^2 e^D=0\,.\label{toda}
\end{equation}
For each ${\cal N}=2$ SCFT described by a Riemann surface with punctures, GM provide a set of boundary conditions which $D$ has to satisfy. The Riemann surface of the gauge theory corresponds to the $x_1\,,\,x_2$ coordinates in M-theory. The Toda equation~(\ref{toda}) appears, in general, to be hard to solve. However, under the assumption of an extra $U(1)$ symmetry in the $x_1\,,\,x_2$ plane it reduces to a Laplace equation~\cite{Ward:1990qt}. General solutions to the Laplace equation with GM boundary conditions were found in~\cite{ReidEdwards:2010qs}, while~\cite{Aharony:2012tz} proposed more general boundary conditions relevant to the gauge/string duality.

In this paper we write down the explicit form of the Bergshoeff-Sezgin-Townsend (BST) action~\cite{Bergshoeff:1987cm} for supermembranes in the GM background in a particular $\kappa$-gauge.~\footnote{Our action, like the GM spacetime, depends implicitly on the function $D$.} The BST action~\cite{Bergshoeff:1987cm} is written in any background and has $\kappa$-symmetry. For a generic background, knowing the spacetime bosonic fields is not sufficient to write down the BST action; one needs to know the full supergeometry. For maximally supersymmetric backgrounds, it was found that the supergeometry is quite simple~\cite{deWit:1998tk,Kallosh:1998qs,Kallosh:1998zx} (see also~\cite{Kallosh:1997qw,Claus:1998ra}) and the action can be written down explicitly~\cite{Kallosh:1998zx,Dall'Agata:1998wz,deWit:1998yu,Claus:1998fh}. For general backgrounds the supergeometry will be quite involved and finding an explicit form of the BST action to all orders in fermions seems a technically difficult problem. In this paper we find a $\kappa$-gauge in which the GM supergeometry simplifies considerably, allowing us to write down explicit expressions for the BST action.

This paper is organised as follows. In section~\ref{sec2} we set our notation, review the superfield formulation of 11-dimensional supergravity, the BST supermembrane action and the GM solution. In section~\ref{sec3} we write down a $\kappa$-gauge and show that in this gauge $W$, the four-form superfield of 11-dimensional supergravity, is ${\cal O}(\theta^0)$ in superspace. This feature of the proposed $\kappa$-gauge simplifies considerably the torsion constraints and allows us to write down explicit expressions for the supervielbein and superconnection, which can be then be inserted into the BST action. 
We then present our conclusions and include appendices on our gamma-matrix and superalgebra conventions.

\section{Setting the notation - a lightning review}\label{sec2}
\setcounter{equation}{0}

In this section we set our notation by reviewing some aspects of 
11-dimensional supergravity, the BST action and the GM solution. Since much of this material is very well known, we will be brief, and refer the reader to the literature for more details.

\subsection{11-dimensional supergravity in superspace}
\label{sec21}

11-dimensional superspace is given by the 43-component vector
\begin{equation}
Z^\Lambda=(X^\mu,\theta^\alpha)\,,
\end{equation}
where $\Lambda=1,\dots,43$, $\mu=0,\dots, 10$, $\alpha=1,\dots,32$ are curved indices. 11-dimensional supergravity~\cite{Cremmer:1978km} is described in superspace by $W_{rstu}$, a 4-form superfield~\cite{Cremmer:1980ru,Brink:1980az}, whose equation of motion is
\begin{equation}
(\Gamma^{rst}D)_aW_{rstu}(X,\theta)=0\,.
\label{Weom}
\end{equation}
Above, $r,s,\dots=0,\dots,10$ and $a,b,\dots=1,\dots,32$ are tangent-space indices.
The lowest superspace component of  $W_{rstu}$ is $F_{rstu}$, the 4-form field-strength of 11-dimensional supergravity. The next two orders of the superfield are the gravitino field-strength and the supersymmetric variation of the gravitino field-strength
\begin{eqnarray}
(D_{a}W_{rstu}(X,\theta))|_{\theta=0}&=&6(\Gamma_{[rs}{\hat D}_t\psi_{u]})_a(X)\,,
\label{wordth}\\
8(D_{a}({\hat D}_{[r}\psi_{s]})_b)|_{\theta=0}&=&
\bigl({\hat R}_{rsmn}(X)\Gamma^{mn}+4
[T_r{}^{t_1t_2t_3t_4}\,,\,T_s{}^{u_1u_2u_3u_4}]
{\hat F}_{t_1t_2t_3t_4}(X){\hat F}_{u_1u_2u_3u_4}(X)
\nonumber \\ && \qquad
+8T_{[s}{}^{t_1t_2t_3t_4}{\hat D}_{r]}{\hat F}_{t_1t_2t_3t_4}(X)\bigr)_{ab}\,,
\label{wordth2}
\end{eqnarray}
where 'hat' denotes supercovariant derivatives, and
\begin{equation}
T^{rstuv}=\frac{1}{144}\Gamma^{rstuv}-\frac{1}{18}\Gamma^{[stu}\eta^{v]r}\,.
\end{equation}
$W$ contains only physical fields of the theory, and the supertorsion and supercurvature can be written exclusively in terms of $W$ and its (first and second order) derivatives~\cite{Cremmer:1980ru,Brink:1980az}
\begin{eqnarray}
T^t_{rs}&=&T^c_{ab}=T^t_{as}=0\,,\nonumber \\
T^r_{ab}&=&\frac{i}{2}(\Gamma^0\Gamma^r)_{ab}\,,
\nonumber \\
T^a_{rs}&=&-\frac{1}{42}(\Gamma^{tu}\Gamma^0)^{ba}D_aW_{rstu}\,,\nonumber \\
T^c_{ar}&=&\frac{1}{2}W_{pstu}(T_r{}^{pstu})^c{}_a\,,\label{torcstr}\\
R_{ab}^{mn}&=&(\Gamma^0 S^{mnt_1t_2t_3t_4})_{ab}
W_{t_1t_2t_3t_4}\,,\nonumber \\
R_{as}{}^{mn}&=&-\frac{i}{42}\bigl(
(\Gamma^0\Gamma^n\Gamma^{rt}\Gamma^0D)_a
W_{rt}{}^m{}_s
-(\Gamma^0\Gamma^m\Gamma^{rt}\Gamma^0D)_a
W_{rt}{}^n{}_s\nonumber \\&&\qquad\qquad
+(\Gamma^0\Gamma^s\Gamma^{rt}\Gamma^0D)_a
W_{rt}{}^{mn}\bigr)\,,\nonumber \\
R_{mn}{}^b{}_c&=&-\frac{1}{21}
(\Gamma^{ns}\Gamma^0)^{ba}D_cD_aW_{rsmn}
-2(T_{[m}{}^{t_1t_2t_3t_4}D_{n]})^b{}_c
\nonumber \\ &&\qquad\qquad
-[T_m{}^{t_1t_2t_3t_4}\,,\,T_n{}^{u_1u_2u_3u_4}]^b{}_c
W_{t_1t_2t_3t_4}W_{u_1u_2u_3u_4}\,.\label{curvcstr}
\end{eqnarray}
The following combination of gamma matrices appears above 
\begin{equation}
S^{t_1t_2}{}^{t_3t_4t_5 t_6}=
\frac{1}{72}\Gamma^{t_1t_2}{}^{t_3t_4t_5t_6}+
\frac{1}{3}\delta^{[t_3}_{t_1}\delta^{[t_4}_{t_2}\Gamma^{t_5t_6]}\,.
\end{equation}
These constraints arise from the identification of general coordinate and local supersymmetry transformations in spacetime with general coordinate transformations in superspace~\cite{Wess:1977fn}. Together with the equation of motion~(\ref{Weom}), they imply the 
Bianchi identities and the field equations of 11-dimensional supergravity.

The supervielbein $E^A_\Lambda$ and affine connection $\Omega_\Lambda^{rs}$ define the supertorsion 
\begin{equation}
T^C_{AB}=(-)^{\Lambda(B+\nu)}E^\Lambda_AE^\nu_B
(D_\Lambda E^c_\nu-(-)^{\Lambda\nu}D_\nu E^C_\Lambda)\,,
\end{equation}
and Lie-algebra valued supercurvature
\begin{equation}
R_{AB}{}^{rs}=(-)^{\Lambda(B+\nu)}E^\Lambda_A 
E^\nu_B\Bigl(\partial_\Lambda\Omega_\nu^{rs}-
(-)^{\Lambda\nu}\partial_\nu\Omega_\Lambda^{rs}
+(\Omega_\Lambda^{rt}\Omega_\nu^{us}
-(-)^{\Lambda\nu}\Omega_\nu^{rt}\Omega_\Lambda^{us})\eta_{tu}
\Bigr)\,.
\end{equation}
The supercurvature $R_{AB}{}^{rs}$ can be extended to a tensor
$R_{AB}{}^{CD}$ by defining
\begin{equation}
R_{AB}{}^{c}{}_d=\frac{1}{4}R_{AB}^{rs}(\Gamma_{rs})^c{}_d\,,\qquad \mbox{and}\qquad
R_{AB}{}^{cs}=R_{AB}{}^{rb}=0\,.
\end{equation}
In principle, one can use these equations, together with equations~(\ref{torcstr}) and~(\ref{curvcstr}), to obtain the supervielbein and affine connection from $W$, order-by-order in $\theta$. In practice, however, this is technically challenging, and explicit expressions do not exist beyond the ${\cal O}(\theta^2)$ terms for general backgrounds.

\subsection{The BST supermembrane action}

Given a supergeometry specified by the supervielbeine $E^A_\Lambda$ and four-form $W_{rstu}$, which 
satisfy equations~(\ref{Weom}),~(\ref{torcstr}) and~(\ref{curvcstr}), Bergshoeff, Sezgin and Townsend postulated the following action for a supermembrane coupled to this supergeometry
\begin{equation}
S_{\mbox{\scriptsize BST}}=\frac{1}{2}\int d^3\sigma
\sqrt{-g}\bigl(g^{ij}E_i^rE_j^s\eta_{rs}-1\bigr)
+\int d^4\sigma W\,,
\label{bst}
\end{equation}
where $\sigma^i$, for $i=1,2,3$, are the membrane world-volume coordinates and 
\begin{equation}
E^A_i=\partial_i Z^\Lambda E^A_\Lambda\,.
\end{equation}
The cosmological constant term ensures that the worldvolume metric satisfies the embedding equation~\cite{Howe:1977hp}
\begin{equation}
g_{ij}=E^a_iE^b_j\eta_{ab}\equiv T_{ij}\,.
\end{equation}
The action~(\ref{bst}) contains the integral of $W=E^{\Lambda_1}E^{\Lambda_2}E^{\Lambda_3}
E^{\Lambda_4}W_{\Lambda_1\Lambda_2\Lambda_3\Lambda_4}$, the pull-back of the four-form superfield onto a four-dimensional manifold whose boundary is the membrane worldvolume. This term is analogous to the formulation of the WZ term of the Green-Schwarz superstring~\cite{Green:1983wt} action as an integral over a three-form~\cite{Henneaux:1984mh}. The BST action is invariant 
under $\kappa$-transformations~\cite{Green:1983wt,Siegel:1983hh} of the 
form~\cite{Bergshoeff:1987cm}
\begin{eqnarray}
\delta_\kappa E^r&=&0\,,\nonumber \\
\delta_\kappa E^a&=&(1+{\bar\Gamma})^a{}_b\kappa^b\,,\nonumber \\
\delta_\kappa g_{ij}&=&2(X_{ij}-g_{ij}X^k{}_k)\,,
\end{eqnarray}
where $\kappa$ is a 32-component Majorana spinor and a world-volume scalar, and
\begin{eqnarray}
\delta_\kappa E^A&=&\delta_\kappa Z^\Lambda E_\Lambda^A\,,
\nonumber \\
{\bar\Gamma}^a{}_b&=&\frac{\sqrt{-g}}{6}\epsilon^{ijk}
E^r_iE^s_jE^t_k(\Gamma_{rst})^a{}_b\,,
\nonumber\\
X_{ij}&=&-\frac{1}{4}\epsilon_i{}^{k_1k_2}E_{k_1}^rE_{k_2}^s
(\Gamma_{rs})_{ab}\,\delta_\kappa E^b \,E^a_j
\nonumber \\ &&
+\frac{1}{2}\kappa^bE_{k_1}^a(\Gamma_t)_{ab}E^{k_1t}
g_{i[j}(T^{k_2}{}_{k_2}T^{k_3}{}_{k_3]}
+\delta^{k_2}{}_{k_2}T^{k_3}{}_{k_3]})+i\leftrightarrow j\,.
\end{eqnarray}

\subsection{The GM solution}

Let us summarise the M-theory solutions discussed by GM~\cite{g1}. In this paper we will not be explicitly interested in the boundary conditions that $D$ satisfies, but rather in the formal expressions of spacetime fields on $D$. As a result, most of the formulas below can be found already (after a suitable change of signature) in~\cite{llm} and~\cite{lm}. The M-theory solution dual to an {\cal N=2} SCFT is
\begin{eqnarray}\label{11d}
ds_{11}^2&=&\kappa^{\frac{2}{3}}e^{2\tilde{\lambda}}\left(4ds^2_{AdS_5}+y^2e^{-6\tilde{\lambda}}d\widetilde{\Omega}_2^2+ds^2_4\right)\nonumber\\
ds^2_4&=&\frac{4}{1-y\partial_yD}(d\chi+v)^2-\frac{\partial_yD}{y}[dy^2+e^D(dx^2_1+dx^2_2)]\nonumber\\
e^{-6\lambda}&=&-\frac{\partial_yD}{y(1-y\partial_y D)}\,,\nonumber \\
v_i&=&\frac{1}{2}\epsilon_{ij}\partial_j D\nonumber\\
G_4&=&\kappa F_2\wedge d\Omega_2\nonumber\\
F_2&=&2(dt+v)\wedge d(y^3e^{-6\tilde{\lambda}})+2y(1-y^2e^{-6\tilde{\lambda}})dv-\partial_ye^Ddx_1\wedge dx_2\,,
\label{GM}
\end{eqnarray}
where $D\equiv D(y,x_1,x_2)$ satisfies the Toda equation~(\ref{toda}). 


The LLM and GM solutions were found by analysing the Killing Spinor Equations (KSE). This method was originally developed for $AdS_5$ backgrounds preserving ${\cal N}=1$ supersymmetry~\cite{Gauntlett:2004zh}. We briefly recall some of the key features here. In 11-dimensional supergravity the supersymmetry variation of the gravitino is
\begin{equation}
\delta_{\varepsilon}\psi_\mu
=\bigl(D_\mu({\hat \omega})
+T_{\mu}{}^{\nu_1\nu_2\nu_3\nu_4}{\hat F}_{\nu_1\nu_2\nu_3\nu_4}
\bigr)\varepsilon\,,
\end{equation}
where
\begin{eqnarray}
{\hat \omega}_{\mu rs}&=&\omega_{\mu rs}
+\frac{i}{2}({\bar\psi}_\mu\gamma_s\psi_r
-{\bar\psi}_\mu\gamma_r\psi_s+{\bar\psi}_s\gamma_\mu\psi_r)\,,\\
{\hat F}_{\mu_1\mu_2\mu_3\mu_4}&=&F_{\mu_1\mu_2\mu_3\mu_4}
-3{\bar\psi}_{[\mu_1}\gamma_{\mu_2\mu_3}\psi_{\mu_4]}\,.
\end{eqnarray}
In general, the Killing spinors $\eta$ of a supersymmetric solution of 11-dimensional supergravity satisfy the KSE, which follow from the vanishing of the supersymmetry variation of the gravitino~\footnote{Since we are only interested in bosonic solutions, we set all fermions on the r.h.s. to zero.} 
\begin{equation}
0=\delta_\eta\psi_m= D_m\eta+T_m{}^{n_1n_2n_3n_4}F_{n_1n_2n_3n_4}\eta\,,
\end{equation}
To write down the LLM Killing spinors, we follow the gamma-matrix conventions of~\cite{llm}
\begin{eqnarray}
\Gamma^{m=0,1,2,3,4}&=&\rho^{m=0,1,2,3,4}\otimes \gamma^7\,,
\qquad
\Gamma^{m=5,6}=1\otimes \sigma^{1,2}\otimes \gamma^5\,,
\qquad\nonumber \\
\Gamma^{m=7,8,9,10}&=&1\otimes 1\otimes \gamma^{1,2,3,4}\,,
\nonumber \\
\gamma^7&=&\sigma^3\otimes\gamma^5\,,\qquad
\gamma^5=\gamma^1\gamma^2\gamma^3\gamma^4\,,
\end{eqnarray}
where $\gamma^{1,2,3,4}$ and $\rho^{0,1,2,3,4}$ are $SO(4)$ and $SO(1,4)$ gamma-matrices, respectively. Above, the $AdS_5$ directions are $m=0,1,2,3,4$ and the $S^2$ directions are $m=5,6$. In this spinor-basis, the Killing spinors of the LLM and GM solutions were found to be~\cite{llm}, 
\begin{equation}
\eta=\psi_{AdS_5}\otimes e^{\lambda/2}\xi\,,
\label{ks11d}
\end{equation}
where $\psi_{AdS_5}$ is the $AdS_5$ Killing spinor (which has 4 complex components) and $\varepsilon$ is
\begin{equation}
\xi=(1-\gamma_5{\hat\gamma})\chi_+\otimes\epsilon
\end{equation}
Above, ${\hat \gamma}\equiv i\gamma^7\gamma^5$, $\chi_+$ is an $S^2$ Killing spinor (which has 2 complex components)~\footnote{We refer the reader to equation~(F.10) of~\cite{llm} for its explicit definition.} and $\epsilon$ is a 
four-component spinor which is given by
\begin{equation}
\epsilon=e^{\zeta \gamma^3/2}{\tilde\epsilon}\,,
\label{4dks}
\end{equation}
where~\footnote{When comparing the above to expressions in Appendix F of~\cite{llm}, the reader should note the following. In~\cite{llm} the authors are interested in a Wick-rotated version of the solution we are reviewing here. As a result, they use superscripts $0,1,2,3$ on the gamma matrices corresponding to the four directions transverse to $S^2$ and $AdS_5$ (or more properly $S^5$ in the Wick rotated setting of~\cite{llm}). In our setting these four dimensions have a Euclidean, rather than Minkowski, signature and it is more natural to label the gamma matrices with a superscript $1,2,3,4$.} $\zeta$ is related to the coordinate $y$ via~\cite{Gauntlett:2004zh}
\begin{equation}
y=e^{3\lambda}\sin\zeta\,,
\label{defzeta}
\end{equation}
and the spinor ${\tilde\epsilon}$ satisfies the projections
\begin{equation}
(1-i\gamma^{123}){\tilde\epsilon}=0\,,\qquad\qquad
(1-i\gamma^{23}){\tilde\epsilon}=0\,.
\end{equation}
Notice that the Killing spinors depend on the $AdS_5$ and $S^2$ coordinates. The dependence on the $S^1$ coordinate is obtained in~\cite{llm}. The Killing spinors also depend on the coordinates $y,x_i$, but only through the specific combination $\zeta=\zeta(y,x_1,x_2)$.
 
As expected, the Killing spinors $\eta$ have 8 complex (or 16 real) independent components, since the GM backgrounds preserve 16 real supersymmetries. We will find it convenient to define the projector ${\tilde\Pi}$ which projects a generic 32-component 11-dimensional Dirac spinor $\Psi$ onto the 16-dimensional Killing spinor $\eta$
\begin{equation}
\eta={\tilde\Pi}\Psi\,.
\end{equation}
The explicit form of ${\tilde\Pi}$ follows from equations~(\ref{ks11d})-(\ref{4dks})
\begin{equation}
{\tilde\Pi}\equiv\frac{1}{8\cos\zeta}\left(1- \Gamma^{5678910}\right)
\left(e^{-i\frac{\zeta}{2}\Gamma^9}(1+i\Gamma^{78})(1+\Gamma^{10})e^{-i\frac{\zeta}{2}\Gamma^9}\right)
\left(1+ \Gamma^{5678910}\right)
\,,
\label{projcomplex}
\end{equation}
and one can easily check that ${\tilde\Pi}^2={\tilde\Pi}$. 
In fact, spinors in 11-dimensional supergravity are Majorana, and a bit of care must be taken when implementing this condition on the Killing spinors $\eta$ which are not Majorana.~\footnote{I am grateful to Dan Waldram for a detailed discussion of this.} Ultimately, projecting onto {\em Majorana} Killing spinors that enter the torsion constraints reviewed above, is not done by the projector ${\tilde \Pi}$, but rather by a related projector $\Pi$ given by
\begin{equation}
\Pi={\tilde\Pi} +B{\tilde\Pi}^*B^{-1}\,,
\end{equation}
where $B$ satisfies
\begin{equation}
(\Gamma^m)^*=B\Gamma^m B^{-1}\,,
\end{equation}
with ${}^*$ denoting complex conjugation.
One can check that $\Pi$ is compatible with the Majorana condition
\begin{equation}
B\Pi=\Pi^* B^*
\end{equation}
which explicitly is
\begin{eqnarray}
\Pi&\equiv&\frac{1}{4\cos\zeta}\left({\bf 1}_{32}- \Gamma^{5678910}\right)
\left(e^{\frac{\zeta}{2}\Gamma^{789}}({\bf 1}_{32}-\Gamma^{10})e^{\frac{\zeta}{2}\Gamma^{789}}\right)\left({\bf 1}_{32}+ \Gamma^{5678910}\right)\nonumber \\
&=&\frac{1}{2 \cos\zeta}\left(\cos\zeta{\bf 1}_{32}-\Gamma^{56789}
-\sin\zeta\Gamma^{5610}\right)
\nonumber\\
&=&
\frac{1}{2\cos\zeta}e^{-\frac{\zeta}{2}\Gamma^{5610}}
\left({\bf 1}_{32}-\Gamma^{56789}\right)e^{-\frac{\zeta}{2}\Gamma^{5610}}
\,.
\label{proj}
\end{eqnarray}
It is easy to check that $\Pi^2=\Pi$ . Since $\Pi$ projects a general spinor $\Psi$ onto a spinor that satisfies the KSE, the supersymmetry variation of $\Pi\Psi$ vanishes
\begin{equation}
\delta_{\varepsilon}(\Pi\psi) = 0\,.
\label{projsusy}
\end{equation}
This identity will play an important role in our choice of $\kappa$ gauge.

\section{The BST action in a suitable $\kappa$-gauge}\label{sec3}
\setcounter{equation}{0}
In order to write down an explicit form of the BST action in a given supergravity background, one first has to know the complete $\theta$ expansion of the superfield $W$. One can then insert this into the supertorsion and supercurvature constraints~(\ref{torcstr}) and~(\ref{curvcstr}) and, in principle, work out the supervielbein order-by-order in $\theta$. A major simplification was found to occur in the maximally supersymmetric backgrounds $AdS_4\times S^7$ and $AdS_7\times S^4$~\cite{Kallosh:1998qs}; it was observed that the superfield $W$ was supercovariant
\begin{equation}
D_AW=0\,.
\label{wsucov}
\end{equation}
This simplifies considerably the constraints~(\ref{torcstr}) and~(\ref{curvcstr}), allowing for a straightforward algebraic derivation of the supervielbein and affine connection~\cite{Kallosh:1998zx}. One can also show~\cite{Claus:1998fh} that this derivation agrees with the superalgebra based expressions for the supervielbein and affine connection~\cite{deWit:1998yu}. Recall that, to show that $W$ is supercovariant for these backgrounds, one need only show that its $O(\theta)$ and $O(\theta^2)$ components are zero~\cite{Kallosh:1998qs}. The former is trivially zero since this is a bosonic background (see equation~(\ref{wordth})), the latter is proportional to the supersymmetry variation of the field-strength of the gravitino (see equation~(\ref{wordth2}))
\begin{equation}
W|_{O(\theta^2)\mbox{term}}\sim 
\delta_\varepsilon({\hat D}_{[\mu}\psi_{\nu]})
={\hat D}_{[\mu}\delta_\varepsilon\psi_{\nu]}\,.
\label{wordth2assusyvar}
\end{equation}
Since these backgrounds are maximally supersymmetric the KSE is satisfied for all spinors, and so
\begin{equation}
\delta_\varepsilon\psi_{\nu}=0\,,
\end{equation}
thus showing that the $O(\theta^2)$ term is zero for the maximally supersymmetric backgrounds. One can confirm this with a straightforward explicit calculation~\cite{Kallosh:1998qs} by plugging in the supergravity solution explicitly into equation~(\ref{wordth2}).

From this argument we see immediately that for backgrounds that are not maximally supersymmetric the $O(\theta^2)$ term has to be non-zero: if it were zero we would conclude that the KSE would be satisfied for all spinors - but that can only be true of the maximally supersymmetric solutions. So it appears that for non-maximally supersymmetric backgrounds extracting the supergeometry from the torsion constraints is a daunting task. However, our goal is not so much the supergeometry, as the BST action. As reviewed in the previous section, the BST action has $\kappa$-symmetry, and our claim is that with the right $\kappa$-gauge choice one can write down an explicit form for the action. Choosing a suitable $\kappa$-gauge has also been useful in writing down explicit  string theory actions in certain backgrounds~\cite{ads4,ads3}. 

Picking a $\kappa$-gauge effectively restricts the superspace coordinates $\theta$ to a 16-dimensional  subspace. In the GM backgrounds there is a natural $\kappa$-gauge choice dictated by the Killing spinors namely we pick the 16 fermions $\theta_\kappa$ that satisfy
\begin{equation}
\Pi\theta_\kappa=\theta_\kappa\,,
\label{kg}
\end{equation}
where the projection $\Pi$ was defined in equation~(\ref{proj}) above. This is the so-called Killing spinor gauge, which has appeared in other settings in the past~\cite{Kallosh:1998nx,Kallosh:1998ji,ads3}. Using equations~(\ref{projsusy}) and~(\ref{wordth2assusyvar}) one can now show that the $O(\theta^2)$ component of the superfield $W$ is zero.~\footnote{The background is bosonic, so the $O(\theta)$ term in the expansion of $W$ is trivially zero; see equation~(\ref{wordth}).} In other words, in the $\kappa$-gauge~(\ref{kg}), $W$ is supercovariant, and the supertorsion and supercurvature constraints simplify considerably
\begin{eqnarray}
T^r&\equiv& d E^r-E^s\Omega_s{}^t=-E^a(\Gamma^0\Gamma^r)_{ab}E^b\,,\nonumber \\
T^a&\equiv& d E^a-\frac{1}{4}\Omega^{st}(\Gamma_{st})^a{}_bE^b=E^r(T_r{}^{s_1s_2s_3s_4})^a{}_b
W_{s_1s_2s_3s_4}E^b\,,\nonumber \\
R^{st}&\equiv&
d\Omega^{rs}-\Omega^r{}_t{}_\wedge\Omega^{ts}
=\frac{1}{2}E^{t_1}E^{t_2}R_{t_1t_2}{}^{rs}
+\frac{1}{2}E^a(\Gamma^0 S^{rst_1t_2t_3t_4})_{ab}
W_{t_1t_2t_3t_4}E^b\,.
\label{thefew}
\end{eqnarray}
These conditions have the same form as the constrains for the maximally supersymmetric backgrounds, albeit now with the fermions restricted to a sixteen-dimensional subspace by the $\kappa$-gauge~(\ref{kg}). By rescaling the $\theta$ coordinates~\cite{Kallosh:1998zx}
\begin{equation}
\theta\rightarrow t\theta\,,
\end{equation}
for some generic c-number $t$, one can arrive at a series of first-order in $t$ differential equations which can be solved given the "initial conditions"~\cite{Kallosh:1998zx}
\begin{equation}
E^a|_{t=0}=0\,,\qquad
E^r|_{t=0}=dX^\mu E_\mu{}^r(X)\,,\qquad
E^a|_{t=0}=dX^\mu\omega_\mu{}^{rs}(X)\,.
\end{equation}
A comprehensive derivation of the solution to this problem is given in~\cite{Kallosh:1998zx,Claus:1998fh} and we simply quote the final result~\footnote{As a result of the Grassmann nature of the fermionic variables all of the sums terminate at a finite order. The order at which they terminate is half of the maximally supersymmetric cases - since here we only have 16 real fermionic degrees of freedom.} 
\begin{eqnarray}
E^a&=&\sum_{n=0}^{8}\frac{1}{(2n+1)!}{\cal M}^n D\theta_\kappa\,,\nonumber \\
E^r&=&dX^\mu e_\mu{}^r(X)+2\sum_{n=0}^{9}\frac{1}{(2n+2)!}\theta_\kappa^a(\Gamma^0\Gamma^r{\cal M}^n)_{ab}
(D\theta_\kappa)^b\,,\nonumber \\
\Omega^{rs}&=&dX^\mu\omega_\mu{}^{rs}(X)-
\sum_{n=0}^{9}\frac{1}{(2n+2)!}\theta_\kappa^a
(\Gamma^0 S^{rst_1t_2t_3t_4}{\cal M}^n)_{ab}
(D\theta_\kappa)^b W_{t_1t_2t_3t_4}\,,
\label{sugeom}
\end{eqnarray}
where
\begin{eqnarray}
{\cal M}_a{}^b&=&2W_{s_1s_2s_3s_4}(\Gamma^0T_r{}^{s_1s_2s_3s_4}\theta_\kappa)_a
(\theta_\kappa\Gamma^0\Gamma^r)^b\nonumber \\
&&-\frac{1}{4}(\Gamma_{rt}\theta_\kappa)_a
W_{s_1s_2s_3s_4}(\theta_\kappa\Gamma^0S^{rts_1s_2s_3s_4})^b\,,\nonumber \\
\theta_\kappa^a&=&=\theta^\alpha E_\alpha{}^a(X)\,,
\end{eqnarray}
and
\begin{equation}
D\theta_\kappa=(d+\frac{1}{4}\omega\cdot\Gamma+E^rT_{r}{}^{s_1s_2s_3s_4}F_{s_1s_2s_3s_4})\theta_\kappa\,.
\end{equation}

The above expression for the supervielbeine can then be inserted into equation~(\ref{bst}) to give the action for a supermembrane in a GM background in the $\kappa$-gauge~(\ref{kg}).~\footnote{We remind the reader that in the $\kappa$-gauge~(\ref{kg}) the super-four-form $W$ does not receive any higher-order $\theta$ corrections.} One may compare the action obtained in this way from the quadratic-order in fermions action~\cite{deWit:1998tk} upon fixing of the $\kappa$-gauge~(\ref{kg}). We have done such a comparison for the Maldacena-Nu\~nez background and found complete agreement up to quadratic order in fermions. Since the expressions are quite lengthy and not very illuminating we do not include them here.

\section{Conclusions}

In this note we have derived an explicit, all-order in fermions, action for supermembranes in GM spacetimes in a particular $\kappa$-gauge. Membranes moving in such spacetimes are conjectured to be dual to ${\cal N}=2$ four-dimensional gauge theories. As such, knowing the dual membrane action is an important step in understanding these gauge/string dualities. 

General solutions to the Toda equation~(\ref{toda}) with GM boundary conditions are not known. If one assumes the presence of a $U(1)$ isometry amongst the $x_i$, the Toda equation reduces to a Laplace equation~\cite{Ward:1990qt} and the boundary value problem can be solved in this case for the GM boundary conditions~\cite{ReidEdwards:2010qs}. Reducing M-theory on this circle, one obtains Type IIA string theory backgrounds which are believed to be the string theory duals of ${\cal N}=2$ four-dimensional gauge theories. In~\cite{Aharony:2012tz}  more general boundary conditions have been proposed, to encode the near-horizon limit of NS5-brane positions. A common feature of all these Type IIA backgrounds is the presence of regions of spacetime in which the dilaton field is large. Under the assumption of a $U(1)$ isometry, the BST supermembrane action has been shown to reduce to a Green-Schwarz action on the KK-reduced spacetime~\cite{Duff:1987bx}. An explicit dictionary exists for re-writing the 11-dimensional supervielbein in terms of 10-dimensional supervielbeine which can then be inserted into the GS actions~\cite{Duff:1987bx}; an explicit discussion of the dilaton couplings can be found in~\cite{Tseytlin:1996hs}. It is straightforward to apply these general formulas to our expressions~(\ref{sugeom}) and in this way obtain $\kappa$-gauge-fixed superstring actions for the backgrounds constructed in~\cite{ReidEdwards:2010qs,Aharony:2012tz}. 

It would be interesting to investigate these superstring actions in more detail, by for example identifying (classically) closed subsectors analogous to the ones found in~\cite{AAT,bsll}. Another interesting question would be to investigate the integrability of such superstring actions. For general solutions of the Toda or Laplace equation we do not expect integrability to be present. However, one may wonder whether certain special solutions lead to integrable string actions. In view of the results~\cite{Gadde:2012rv}, it would also be interesting to see under what circumstances integrable sub-sectors exist for these string actions, for example in analogy with the ones found in~\cite{AAT,bsll}. Given recent progress in obtaining quartic order in fermions string actions~\cite{Wulff:2013kga}, it would be interesting to apply these results to match our expressions. Finally, we note that the methods used here to obtain supergeometries and corresponding membrane and string actions in a suitable $\kappa$-gauge should be applicable to other backgrounds preserving 16 real supersymmetries. So, for example, if one were able to circumvent the no-go result of~\cite{Colgain:2011hb} and find Type IIB solutions with ${\cal N}=2$ supersymmetry and an $AdS_5$ factor, it would then be possible to obtain the relevant string actions using methods similar to the ones we have described here.

\section*{Acknowledgements}

I would like to thank Jerome Gauntlett, Chris Hull,  Dario Martelli, Carlos Nu\~nez, James Sparks, Kelly Stelle, Arkady Tseytlin, Dan Waldram, Kostya Zarembo for valuable discussions and for sharing their insights with me. I am greatful to the Tata Insititute of Fundamental Research in Mumbai, Kavli Institute for Theoretical Physics in Santa Barbara and the Centro de Ciencias de Benasque Pedro Pascual in Benasque  for hospitality where during parts of this project. I gratefully acknowledge funding support from an EPSRC Advanced Fellowship and an STFC Consolidated Grant "Theoretical Physics at City University" ST/J00037X/1.

\appendix
\section{11-dimensional gamma matrices}

We will use the following conventions for the $32\times 32$ gamma matrices.
\begin{eqnarray}
\Gamma^0&=&i 
\gamma^5\otimes \gamma^7\,,
\qquad
\Gamma^1=
\gamma^1\otimes \gamma^7\,,
\qquad
\Gamma^2= 
\gamma^2\otimes \gamma^7\,,
\qquad
\nonumber \\
\Gamma^3&=&
\gamma^3\otimes \gamma^7\,,
\qquad
\Gamma^4=
\gamma^4\otimes \gamma^7\,,
\nonumber \\
\Gamma^{m=5,6}&=&{\bf 1}_4\otimes \sigma^{1,2}\otimes \gamma^5\,,
\qquad
\Gamma^{m=7,8,9,10}={\bf 1}_8\otimes \gamma^{1,2,3,4}\,,
\nonumber \\
\gamma^7&=&\sigma^3\otimes\gamma^5\,,\qquad
\gamma^5=\gamma^1\gamma^2\gamma^3\gamma^4\,.
\end{eqnarray}
It will be particularly useful for us to consider 
\begin{eqnarray}
\gamma^1&=&\sigma^2\otimes \sigma^1\,,\qquad
\gamma^2=\sigma^2\otimes \sigma^2\,,\qquad
\gamma^3=\sigma^2\otimes \sigma^3\,,\qquad
\nonumber \\
\gamma^4&=&\sigma^1\otimes {\bf 1}_2\,,\qquad
\gamma^5 =\sigma^3 \otimes {\bf 1}_2 \,.
\label{4dgammaB}
\end{eqnarray}
We define the matrices $B$, $C$ and $T$ in the conventional way
\begin{equation}
(\Gamma^m)^*=B\Gamma^m B^{-1}\,,\qquad
(\Gamma^m)^t=-T\Gamma^m T^{-1}\,,\qquad
(\Gamma^m)^\dagger=-C\Gamma^m C^{-1}\,,
\end{equation}
where ${}^*\,,\,{}^t\,,\,{}^\dagger\,,\,$ are complex conjugation, transpose and hermitian conjugation, respectively. The matrices defined above satisfy
\begin{equation}
BB^*={\bf 1}_{32}\,,\qquad T T^t=-{\bf 1}_{32}\,,\qquad
C C^\dagger = {\bf 1}_{32}\,.
\end{equation}
Note in particular that the first of the above identities allows one to impose the Majorana condition.
In the explicit basis~(\ref{4dgammaB}) above these matrices are
\begin{equation}
B=\Gamma^2\Gamma^4\Gamma^5\Gamma^8\Gamma^{10}\,,\qquad
C=\Gamma^0\,,\qquad T=B\Gamma^0\,,
\end{equation}
In our explicit basis we have
\begin{equation}
T=-i {\bf 1}_2\otimes \epsilon\otimes \epsilon\otimes \sigma^3\otimes \epsilon\,.
\end{equation}


\section{Some superalgebras}

We collect here some information about the commutation relations of superalgebras $OSp(1,7|4)$, $OSp(1,7|2)$ and $SU(2,2|2)$.

\subsection{The $OSp(1,7|4)$ algebra}

Recall that the $OSp(1,7|4)$ super-algebra relations are
\begin{eqnarray}
\left[P^A\,,\,P^B\right]&=&J^{AB}\,,\qquad
\left[P^A\,,\,J^{BC}\right]=\eta^{AB}P^C-\eta^{AC}P^B\,,\\
\left[J^{AB}\,,\,J^{CD}\right]&=&\eta^{BC}J^{AD}\pm\mbox{3 terms}\,,\qquad \left[I^{IJ}\,,\,I^{KL}\right]=\delta^{JK}I^{IL}\pm\mbox{3 terms}\,,\\
\left[Q_{{\hat a}{\hat \alpha}}\,,\,J^{AB}\right]&=&\frac{1}{2}
Q_{{\hat b}{\hat \alpha}}(\rho^{AB})^{\hat b}{}_{\hat a} \,,\qquad
\left[Q_{{\hat a}{\hat \alpha}}\,,\,P^A\right]=\frac{1}{2}
Q_{{\hat b}{\hat \alpha}}(\rho^A)^{\hat b}{}_{\hat a} \,,\\
\left[Q_{{\hat a}{\hat \alpha}}\,,\,I^{IJ}\right]&=&-\frac{1}{2}
Q_{{\hat a}{\hat\beta}}(\gamma^{IJ})^{\hat\beta}{}_{\hat\alpha} \,,\qquad\\
\left\{Q_{{\hat a}{\hat\alpha}}\,,\,Q_{{\hat b}{\hat\beta}}\right\}&=&i(t\rho^{AB})_{{\hat a}{\hat b}}{\hat t}_{{\hat \alpha}{\hat \beta}}J^{AB}
-2i(t\rho^A )_{{\hat a}{\hat b}}{\hat t}_{{\hat\alpha}{\hat\beta}}P^A
-2it_{{\hat a}{\hat b}}({\hat t}\gamma^{IJ} )_{{\hat\alpha}{\hat\beta}}I^{IJ}\,.
\label{fermcomm}
\end{eqnarray}
Above, $\eta^{AB}=\mbox{diag}(-1\,,\,1\,,\,1\,,\,1\,,\,1\,,\,1\,,\,1)$, ${\hat a}=1\,\dots\,8$ is a Dirac $SO(1,7)$ spinor index, ${\hat\alpha}=1\,\dots\,4$ is a Dirac $SO(5)$ spinor index, $A,B=0,\dots,6$ is a $SO(1,7)$ vector index and $I,J=1,\dots,5$ is a $SO(5)$ vector index. Above
\begin{equation}
\gamma^{IJ}=\frac{1}{2}(\gamma^I\gamma^J-\gamma^J\gamma^I)\,,
\end{equation}
with an explicit basis of $\gamma^I$ given in equation~(\ref{4dgammaB}) above. The matrices $\rho^A$
satisfy
\begin{equation}
\left\{\rho^A\,,\,\rho^B\right\}=2\eta^{AB}\,,
\end{equation}
and
\begin{equation}
\rho^{AB}=\frac{1}{2}(\rho^A\rho^B-\rho^B\rho^A)\,.
\end{equation}
An explicit basis is given by
\begin{eqnarray}
\rho^0&=&i\gamma^5\otimes \sigma^3\,,\qquad
\rho^1=\gamma^1\otimes \sigma^3\,,\qquad
\rho^2=\gamma^2\otimes \sigma^3\,,\nonumber \\
\rho^3&=&\gamma^3\otimes \sigma^3\,,\qquad
\rho^4=\gamma^4\otimes \sigma^3\,,\qquad
\rho^{5,6}={\bf 1}_4\otimes \sigma^{1,2}\,.
\end{eqnarray}
The matrices $t$ and ${\hat t}$ are defined via
\begin{equation}
(\rho^A)^t=-t\rho^At^{-1}\,,\qquad
(\gamma^I)^t={\hat t}\gamma^I{\hat t}^{-1}\,.
\end{equation}
The matrices $t\rho^A$, $t\rho^{AB}$ and ${\hat t}$ are anti-symmetric, while $t$ and ${\hat t}\gamma^{IJ}$ are symmetric.
In the explicit basis we are using $t$ and ${\hat t}$ can be written as
\begin{equation}
{\hat t}=\gamma^1\gamma^3\,,\qquad
t=\rho^1\rho^3\rho^6={\hat t}\otimes \sigma^2\,.
\label{tintosp174}
\end{equation}
The supercharges satisfy the reality condition
\begin{equation}
(Q_{{\hat a}{\hat\alpha}})^*=b_{\hat a}{}^{\hat b}{\hat b}_{\hat\alpha}{}^{\hat\beta}Q_{{\hat b}{\hat\beta}}\,,
\label{sualgreal}
\end{equation}
where $b$ and ${\hat b}$ are defined via
\begin{equation}
(\rho^A)^*=b\rho^A b^{-1}\,,\qquad
(\gamma^I)^*={\hat b}\gamma^I {\hat b}^{-1}\,,
\end{equation}
and in our basis
\begin{equation}
b=\rho^0t\,,\qquad
{\hat b}={\hat t}\,.
\label{bintosp174}
\end{equation}
In the basis we are using we have the following useful identity
\begin{equation}
B=b\otimes {\hat b}\,.
\end{equation}
The reality condition~(\ref{sualgreal}) is consistent since
\begin{equation}
b^*b=-{\bf 1}_8\,,\qquad {\hat b}^*{\hat b}=-{\bf 1}_4\,.
\end{equation}
It is sometimes useful to write the $I^{IJ}$ as 
\begin{equation}
P^i\equiv I^{i5}\,,\qquad\mbox{ and }\qquad I^{ij}\,,
\end{equation}
where $i,j=1,2,3,4$. Some useful identities involving the $\rho$ and $\gamma$ matrices are
\begin{eqnarray}
2(\rho_A)^{{\hat d}}{}_{{\hat a}}(t\rho^A)_{{\hat b}{\hat c}}-(\rho_{AB})^{{\hat d}}{}_{{\hat a}}(t\rho^{AB})_{{\hat b}{\hat c}}&=&8t_{{\hat a}{\hat b}}\delta^{{\hat d}}_{{\hat c}}-8t_{{\hat a}{\hat c}}\delta^{{\hat d}}_{{\hat b}}\,,
\label{rhof}
\\
(\gamma_{IJ})^{{\hat \delta}}{}_{{\hat\alpha}}({\hat t}\gamma^{IJ})_{{\hat\beta}{\hat\gamma}}&=&4{\hat t}_{{\hat\alpha}{\hat \beta}}\delta^{{\hat \delta}}_{{\hat \gamma}}+4{\hat t}_{{\hat\alpha}{\hat\gamma}}\delta^{{\hat\delta}}_{{\hat\beta}}\,,
\label{gamaf}
\end{eqnarray}
Using these identities one can, for example, check that the Jacobi identity $\left[Q\,,\,\left\{Q\,,\,Q\right\}\right]+\dots$ is satisfied.

\subsection{The $SU(2,2|2)$ algebra}

Recall that the $SU(2,2|2)$ super-algebra relations are
\begin{eqnarray}
\label{su222}
\left[{\tilde P}^\tA\,,\,{\tilde P}^\tB\right]&=&{\tilde J}^{\tA\tB}\,,\qquad
\left[{\tilde P}^\tA\,,\,{\tilde J}^{\tB\tC}\right]=\eta^{\tA\tB}{\tilde P}^\tC-\eta^{\tA\tC}{\tilde P}^\tB\,,\\
\left[{\tilde J}^{\tA\tB}\,,\,{\tilde J}^{\tC\tD}\right]&=&\eta^{\tB\tC}J^{\tA\tD}\pm\mbox{3 terms}\,,\qquad \left[M^{ij}\,,\,M^{kl}\right]=\delta^{jk}M^{il}\pm\mbox{3 terms}\,,\\
\left[Q^X_{a\alpha}\,,\,{\tilde J}^{\tA\tB}\right]&=&-\frac{1}{2}
Q^X_{b\alpha}({\tilde\gamma}^{\tA\tB})^{b}{}_{a} \,,\qquad
\left[Q^X_{a\alpha}\,,\,{\tilde P}^\tA\right]=\frac{1}{2}
\epsilon^{XY}Q^Y_{b\alpha}({\tilde\gamma}^A)^{b}{}_{a} \,,\\
\left[Q^X_{a\alpha}\,,\,M^{ij}\right]&=&-\frac{1}{2}
Q^X_{a\beta}(\sigma^{ij})^{\beta}{}_{\alpha} \,,\qquad
\left[Q^X_{a\alpha}\,,\,P\right]=\frac{1}{2}\epsilon^{XY}
Q^Y_{a\alpha} \,,\\
\left\{Q^X_{a\alpha}\,,\,Q^Y_{b\beta}\right\}&=&i\epsilon^{XY}({\tilde t}{\tilde\gamma}^{\tA\tB})_{ab}\epsilon_{\alpha\beta}{\tilde J}^{\tA\tB}
-2i\delta^{XY}({\tilde t}{\tilde\gamma}^\tA )_{ab}\epsilon_{\alpha\beta}{\tilde P}^\tB
\nonumber \\ &&
-2i\epsilon^{XY}{\tilde t}_{ab}(\epsilon\sigma^{ij} )_{\alpha\beta}M^{ij}
+2i\delta^{XY}{\tilde t}_{ab}\epsilon_{\alpha\beta}P\,.
\label{su22fermcomm}
\end{eqnarray}
Above, $\eta^{\tA\tB}=\mbox{diag}(-1\,,\,1\,,\,1\,,\,1\,,\,1)$, $a=1\,\dots\,4$ is a Dirac $SO(1,4)$ spinor index, $\alpha=1\,,\,2$ is a Dirac $SO(3)$ spinor index, $\tA,\tB=0,\dots,4$ is a $SO(1,4)$ vector index and $i,j=1,2,3$ is a $SO(3)$ vector index. Above, the ${\tilde\gamma}$ matrices are conventional $SO(1,4)$ gamma matrices which we will take to be
\begin{equation}
{\tilde\gamma}^0\equiv i\gamma^5\,,\qquad
{\tilde\gamma}^1\equiv \gamma^1\,,\qquad
{\tilde\gamma}^2\equiv \gamma^2\,,\qquad
{\tilde\gamma}^3\equiv \gamma^3\,,\qquad
{\tilde\gamma}^4\equiv \gamma^4\,,
\label{41dgamma}
\end{equation}
while ${\tilde t}$ is defined as
\begin{equation}
({\tilde\gamma}^\tA)^t={\tilde t}{\tilde\gamma}^\tA {\tilde t}^{-1}\,,
\end{equation}
and in our basis is
\begin{equation}
{\tilde t}={\tilde\gamma}^1{\tilde\gamma}^3=-{\bf 1}_2\otimes\epsilon\,.
\label{41dinter}
\end{equation}
In fact, ${\tilde t}$ is
We also define
\begin{equation}
{\tilde\gamma}^{\tA\tB}=\frac{1}{2}({\tilde\gamma}^\tA{\tilde\gamma}^\tB
-{\tilde\gamma}^\tB{\tilde\gamma}^\tA)\,.
\end{equation}
The ${\tilde\gamma}$ matrices satisfy the following identities
\begin{eqnarray}
({\tilde\gamma}_\tA)^d{}_a({\tilde t}{\tilde\gamma}^\tA)_{bc}&=&-{\tilde t}_{bc}\delta^d_a-2{\tilde t}_{ab}\delta^d_c+2{\tilde t}_{ac}\delta^d_b\,,
\\
({\tilde\gamma}_{\tA\tB})^d{}_a({\tilde t}{\tilde\gamma}^{\tA\tB})_{bc}&=&4{\tilde t}_{ab}\delta^d_c+4{\tilde t}_{ac}\delta^d_b\,,
\label{tgamaf}
\end{eqnarray}
where the indices $\tA,\tB$ are lowered with the Minkowski metric $\eta_{\tA\tB}$. Using these identities, as well as equation
\begin{equation}
(\sigma_{IJ})^{\delta}{}_{\alpha}(t\rho^{IJ})_{\beta\gamma}=2\epsilon_{\alpha\beta}\delta^{\delta}_{\gamma}
+2\epsilon_{\alpha\gamma}\delta^{\delta}_{\beta}\,,
\label{sigf}
\end{equation}
one can check that the Jacobi identity $\left[Q\,,\,\left\{Q\,,\,Q\right\}\right]+\dots$ is satisfied.
The supercharges satisfy the reality condition
\begin{equation}
(Q_{a\alpha})^*={\tilde b}_a{}^b
\epsilon_\alpha{}^\beta Q_{b\beta}\,,
\label{su222real}
\end{equation}
where ${\tilde b}$ is defined by
\begin{equation}
({\tilde\gamma}^\tA)^*=-{\tilde b}{\tilde\gamma}^\tA {\tilde b}^{-1}\,.
\end{equation}
In our basis we can take
\begin{equation}
{\tilde b}=-{\tilde\gamma}^3{\tilde\gamma}^4\,.
\end{equation}
The reality condition~(\ref{su222real}) is consistent since
\begin{equation}
{\tilde b}^*{\tilde b}=-{\bf 1}_{4}\,,\qquad \epsilon^*\epsilon=-{\bf 1}_{2}\,.
\end{equation}

\section{Projecting the $OSp(1,7|4)$ super-algebra}

Let us consider a number of projectors on the spinors $Q$ and work out what the resulting sub-algebra of $OSp(1,7|4)$ is.

\subsection{LLM Projections}

Let us decompose the $OSp(1,7|4)$ bispinor $Q_{{\hat a}{\hat\alpha}}$ in the LLM fashion as
\begin{equation}
Q_{{\hat a}{\hat\alpha}}=Q_{aa'\alpha'\alpha}\,,
\label{fermdec}
\end{equation}
where the subscript $a=0,\dots,4$ is an $SO(1,4)$ Dirac spinor index, $a'=1,2$ and $\alpha'=1,2$. The LLM Killing spinors can be constructed by a projection operator that acts only on the $a'$ and $\alpha'$ indices. It also is dressed by an extra exponential factor. For simplicity we first consider projections without such an exponential factor.

\subsubsection{LLM projections without exponential dressing phase}
%
Consider the projection
\begin{equation}
\label{pi4}
\Pi_\pm^{(4)}\equiv\frac{1}{2}({\bf 1}_{32}\pm\Gamma^{56789})
=\frac{1}{2}({\bf 1}_{32}\pm{\bf 1}_4\otimes\sigma^3\otimes \sigma^2\otimes {\bf 1}_2)\,.
\end{equation}
This projection is compatible with the Majorana condition, and it projects out half of the fermionic supercharges. The remaining charges can be written as
\begin{eqnarray}
{\tilde Q}^{X=1}_{a\alpha}&\equiv&\frac{1}{2}\left(Q_{a,a'=1,\alpha'=1,\alpha}
+i Q_{a,a'=1,\alpha'=2,\alpha}+iQ_{a,a'=2,\alpha'=1,\alpha}+Q_{a,a'=2,\alpha'=2,\alpha}\right)\,,
\nonumber \\
{\tilde Q}^{X=2}_{a\alpha}&\equiv&\frac{1}{2}\left(i Q_{a,a'=1,\alpha'=1,\alpha}
- Q_{a,a'=1,\alpha'=2,\alpha}+Q_{a,a'=2,\alpha'=1,\alpha}-iQ_{a,a'=2,\alpha'=2,\alpha}\right)\,.
\end{eqnarray}
We have written things in this suggestive way, since the ${\tilde Q}$ will turn out to satisfy the $SU(2,2|2)$ algebra. To see this it is useful to decompose the spinors in equation~(\ref{fermcomm}) as in equation~(\ref{fermdec})
\begin{eqnarray}
\left\{Q_{a\alpha}\,,\,Q_{b\beta}\right\}&=&
-({\tilde t}{\tilde\gamma}^{\tA\tB})_{ab}
\epsilon_{a'b'}\delta_{\alpha'\beta'}\epsilon_{\alpha\beta}J^{\tA\tB}
+2({\tilde t}{\tilde\gamma}^{\tA})_{ab}
\delta_{a'b'}\delta_{\alpha'\beta'}\epsilon_{\alpha\beta}J^{\tA 5}
+2i({\tilde t}{\tilde\gamma}^{\tA})_{ab}
\sigma^3_{a'b'}\delta_{\alpha'\beta'}\epsilon_{\alpha\beta}J^{\tA 6}
\nonumber \\&&
-2({\tilde t}{\tilde\gamma}^\tA )_{ab}
\sigma^1_{a'b'}\delta_{\alpha'\beta'}\epsilon_{\alpha\beta}P^\tA
+2{\tilde t}_{ab}
\sigma^3_{a'b'}\delta_{\alpha'\beta'}\epsilon_{\alpha\beta}P^5
+2i{\tilde t}_{ab}
\delta^1_{a'b'}\delta_{\alpha'\beta'}\epsilon_{\alpha\beta}P^6
\nonumber \\
&&
+2i{\tilde t}_{ab}
\sigma^1_{a'b'}\delta_{\alpha'\beta'}\epsilon_{\alpha\beta}J^{56}
+4{\tilde t}_{ab}
\sigma^2_{a'b'}\sigma^2_{\alpha'\beta'}\epsilon_{\alpha\beta}
I^{45}
\nonumber \\&&
+2{\tilde t}_{ab}
\epsilon_{a'b'}\delta_{\alpha'\beta'}(\epsilon\sigma^{ij})_{\alpha\beta}I^{ij}
-4{\tilde t}_{ab}
\sigma^2_{a'b'}\sigma^3_{\alpha'\beta'}(\epsilon\sigma^i)_{\alpha\beta}I^{4j}
+4{\tilde t}_{ab}
\sigma^2_{a'b'}\sigma^1_{\alpha'\beta'}(\epsilon\sigma^i)_{\alpha\beta}I^{5j}
\,.
\nonumber \\
\label{decomposp172su222}
\end{eqnarray}
Above, by a slight abuse of notation, we decompose the superscripts $A=(\tA,5,6)$ and $I=(i,4,5)$, where $\tA=0,1,\dots,4$ and $i=1,2,3$.  The projection $\Pi^{(4)}_+$ acts only on the primed subscripts, and given the above decomposition it is easy to show that
\begin{eqnarray}
\left\{{\tilde Q}^X_{a\alpha}\,,\,{\tilde Q}^Y_{b\beta}\right\}
&=&-\epsilon^{XY}({\tilde t}{\tilde\gamma}^{\tA\tB})_{ab}\epsilon_{\alpha\beta}{\tilde J}^{\tA\tB}-2i\delta^{XY}({\tilde t}{\tilde\gamma}^{\tA})_{ab}\epsilon_{\alpha\beta}{\tilde P}^{\tA}
+2\epsilon^{XY}{\tilde t}_{ab}(\epsilon\sigma^{ij})_{\alpha\beta}M^{ij}\nonumber \\
&&- 2\delta^{XY}{\tilde t}_{ab}\epsilon_{\alpha\beta}P\,,
\label{su22projqq}
\end{eqnarray}
where
\begin{eqnarray}
P\equiv J^{56}+2I^{45}\,,\qquad
M^{ij}\equiv I^{ij}\,,\qquad
{\tilde J}^{\tA\tB}\equiv J^{\tA\tB}\,,\qquad
{\tilde P}^{\tA}\equiv P^{\tA}\,.
\end{eqnarray}

\subsubsection{LLM projections with exponential dressing phase}

 Consider now the full LLM projector
\begin{eqnarray}
\Pi&=&
\frac{1}{2}e^{-\frac{\zeta}{2}\Gamma^{567910}}
\left({\bf 1}_{32}+\Gamma^{56789}\right)e^{\frac{\zeta}{2}\Gamma^{567910}}\nonumber \\ &=&
\frac{1}{2}e^{-\frac{\zeta}{2}{\bf 1}_4\otimes\sigma^3\otimes\sigma^1\otimes\sigma^2}
\left({\bf 1}_{32}+{\bf 1}_4\otimes\sigma^3\otimes\sigma^2\otimes{\bf 1}_2\right)e^{\frac{\zeta}{2}{\bf 1}_4\otimes\sigma^3\otimes\sigma^1\otimes\sigma^2}
\end{eqnarray}
as given in equation~(\ref{proj}).
The supercharges preserved by the projection can be written as
\begin{eqnarray}
{\tilde Q}^{X=1}_{a\alpha}&\equiv&\frac{\cosh(\zeta/2)}{2\sqrt{\cosh\zeta}}\left(Q_{a,a'=1,\alpha'=1,\alpha}
+i Q_{a,a'=1,\alpha'=2,\alpha}+iQ_{a,a'=2,\alpha'=1,\alpha}+Q_{a,a'=2,\alpha'=2,\alpha}\right)
\nonumber \\&&-i\,
\frac{\sinh(\zeta/2)}{2\sqrt{\cosh\zeta}}\epsilon_{\alpha\beta}\left(Q_{a,a'=1,\alpha'=2,\beta}
+i Q_{a,a'=1,\alpha'=1,\beta}-iQ_{a,a'=2,\alpha'=2,\beta}-Q_{a,a'=2,\alpha'=1,\beta}\right)\,,
\nonumber \\
{\tilde Q}^{X=2}_{a\alpha}&\equiv&\frac{\cosh(\zeta/2)}{2\sqrt{\cosh\zeta}}\left(i Q_{a,a'=1,\alpha'=1,\alpha}
- Q_{a,a'=1,\alpha'=2,\alpha}+Q_{a,a'=2,\alpha'=1,\alpha}-iQ_{a,a'=2,\alpha'=2,\alpha}\right)\nonumber \\
&&-i\,\frac{\sinh(\zeta/2)}{2\sqrt{\cosh\zeta}}\epsilon_{\alpha\beta}\left(i Q_{a,a'=1,\alpha'=2,\beta}
- Q_{a,a'=1,\alpha'=1,\beta}-Q_{a,a'=2,\alpha'=2,\beta}+iQ_{a,a'=2,\alpha'=1,\beta}\right)
\,.\nonumber \\
\label{fullllmprojsucharge}
\end{eqnarray}
Using the decomposition~(\ref{decomposp172su222}) we can show that the supercharges given in equation~(\ref{fullllmprojsucharge})
satisfy equation~(\ref{su22projqq}) with the identification
\begin{eqnarray}
P&\equiv&J^{56}+2I^{45}\mbox{sech}\,\zeta+2I^{25}\tanh\zeta\,,\qquad
{\tilde J}^{\tA\tB}\equiv J^{\tA\tB}\,,\qquad
{\tilde P}^{\tA}\equiv P^{\tA}\,,
\nonumber \\
M^{12}&\equiv&\mbox{sech}\,\zeta I^{12}+\tanh\zeta I^{41}\,,
\qquad
M^{23}\equiv\mbox{sech}\,\zeta I^{23}-\tanh\zeta I^{43}\,,
\qquad
M^{13}\equiv I^{13}.
\end{eqnarray}

\end{document}